\documentstyle[twocolumn,eqsecnum,aps]{revtex}
\begin{document}
\draft
%\preprint{HEP/123-qed}
\title{On the Possible Violation of Sum Rules for
Higher-Twist Parton Distributions}
\author{M. Burkardt}
\address{
Institute for Nuclear Theory\\
University of Washington, Box 351550\\
Seattle, WA 98195-1550}
%\date{\today}
\maketitle
\begin{abstract}
Precise measurements of polarized electro-production and
Drell-Yan in the deep inelastic limit will soon provide
first information on the higher twist parton distributions
$g_T(x)$ and $h_L(x)$.
Sum rules for higher-twist structure
functions are
only valid provided the corresponding Compton amplitudes
satisfy un-subtracted dispersion relations.
Subtracted dispersion relations have to be used when
the (real part of the) forward scattering amplitudes
does not fall off rapidly enough for $\nu \rightarrow
\infty$ (fixed $Q^2$). Formally, such subtractions lead to
$\delta$-functions at the origin in the parton distributions, which are not
accessible to experiment, and the integral over the data
fails to satisfy the sum rule.
The $\delta$-functions in the parton distributions can be identified
with the zero-modes that appear in light-front quantization.
An explicit {\it infinite momentum boost} identifies
these soft quark modes with low momentum contributions
to the self-energy.
\end{abstract}
%\pacs{Valid PACS appear here.
%{\tt$\backslash$\string pacs\{\}} should always be input,
%even if empty.}

\narrowtext

\section{Introduction}
\label{sec:intro}
With the advent of precise deep inelastic scattering (DIS) experiments
it will soon become possible to measure
the proton's polarized higher-twist structure functions
$g_T(x)$ and $h_L(x)$. $g_T(x) \equiv g_2(x)-g_1(x)$ is obtained
in polarized electro-production \cite{hey}, where one finds for
the parallel-anti-parallel asymmetry
\begin{equation}
\frac{d\sigma^{\uparrow \uparrow}}{dq^2dE^\prime}-
\frac{d\sigma^{\uparrow \downarrow}}{dq^2dE^\prime}
= \frac{4\pi \alpha}{E^2 q^2}\left[
\left( E + E^\prime \cos \theta \right) M G_1 + q^2 G_2\right],
\end{equation}
where $E, E^\prime$ are the lab energy of the initial/final lepton,
$\uparrow$, $\downarrow$ denote nucleon/lepton felicities and
$\theta$ is the lab angle of the final lepton.
In the parton model, $\nu G_1$ scales while only $\nu^2 G_2$
has a finite scaling limit as $Q^2 \equiv -q^2 \rightarrow \infty$,
$\nu \equiv E-E^\prime \rightarrow \infty$ and
$x_{Bj}\equiv Q^2/2M\nu$ fixed (the Bjorken limit). Using
the operator product expansion one finds
(modulo logarithmic scaling violations)
$M^2\nu G_1 \stackrel{Bj}{\longrightarrow} \sum_q
e_q^2g^q_1(x_{Bj})+g^q_1(-x_{Bj})$ and
$M \nu^2 G_2 \stackrel{Bj}{\longrightarrow} \sum_q e_q^2
g^q_2(x_{Bj})+g^q_2(-x_{Bj})$,\footnote{$g^q_i(-x)$ is sometimes
denoted $g^{\bar{q}}_i(x)$ in the literature.}
where the parton distributions $g^q_1$ and $g^q_2$ are
defined
through the light cone correlations\cite{jaji}\\
\vspace{1.cm}
\begin{eqnarray}
& &\int \frac{d \lambda}{2\pi} e^{i\lambda x}
\langle PS|\bar{q}(0)\gamma^\mu\gamma_5q (\lambda n)|PS\rangle
\label{eq:g123}
\\
&=&2 \left[g^q_1(x) p^\mu (S \cdot n)
+ g^q_T(x)S_T^\mu + M^2 g^q_3(x) n^\mu (S\cdot n)\right],
\nonumber
\end{eqnarray}
where $n^2=p^2=0$, $n^+=p^-=0$, $P^\mu=p^\mu+\frac{M^2}{2}n^\mu$,
$S^\mu=p^\mu (S\cdot n)+ n^\mu (S\cdot p)+ S_T^\mu$ and
$g^q_T(x)=g^q_1(x)+g^q_2(x)$.
The different scaling behavior of $G_1$ and $G_2$ as
$p^+\rightarrow \infty$ (Breit frame) is reflected in the coefficients
in Eq.(\ref{eq:g123}), where the coefficient of $g^q_2$ is independent of
$p^+$, while the contribution from $g^q_1$ in Eq.(\ref{eq:g123}) grows linearly
with $p^+$. This is typical for operators of different twist.
Clearly, the non-leading role played by $G_2$ makes it rather
difficult to separate them from the leading twist term $G_1$ (and
from $1/Q^2$ corrections to $G_1$).
The HERMES experiment at DESY will be the first attempt for a
precise determination of $g^q_2(x)$ \cite{hermes}, which has
received special attention due to the existence of the
Burkhardt-Cottingham sum rule \cite{bc}
\begin{equation}
\int_{-1}^1 dx g^q_2(x)\equiv \int_{0}^1 dx \left[
g^q_2(x)+g^{\bar{q}}_2(x)\right]=0,
\label{eq:bc0}
\end{equation}
which seems to follow easily from the OPE result (\ref{eq:g123})
\cite{nag92,mulders}.
To see this, one can integrate Eq.(\ref{eq:g123}) over $x$
[using the fact that the r.h.s. has support only for $x \in (-1;1)$], where
we find
\begin{eqnarray}
& &\langle PS|\bar{q}(0)\gamma^\mu\gamma_5q (0)|PS\rangle =
\label{eq:g123sr}
\\
&2\int_{-1}^1&
\left[g^q_1(x) p^\mu (S \cdot n)
+ g^q_T(x)S_T^\mu + M^2 g^q_3(x) n^\mu (S\cdot n)\right].
\nonumber
\end{eqnarray}
from rotational invariance it follows that the l.h.s. in Eq.(\ref{eq:g123sr})
is proportional to the spin vector $S_\mu$ and thus the $g^q_i(x)$ must
satisfy
\begin{eqnarray}
\int_{-1}^1 dx g^q_1(x)&=&\int_{-1}^1dx g^q_T(x)
\label{eq:bc}
\nonumber\\
\int_{-1}^1dx g^q_1(x)&=&2\int_{-1}^1dx g^q_3(x).
\label{eq:g3sr}
\end{eqnarray}
The first of these conditions (\ref{eq:bc}) is abovementioned
Burkhardt-Cottingham sum-rule, while the second is a
(probably useless) sum-rule for $g_3$.\footnote{Useless, because
no experiment is known that could access $g_3$ with
sufficient precision and because Regge arguments suggest that
the sum rule is highly divergent.}

The way I presented the derivation of these sum rules, it
seems that they are mere consequences of rotational invariance
and hence there should be no question about their validity in QCD.
However, there are several subtleties if one wants to turn
these sum rules for the $g_i$'s (the light-cone correlators)
into sum rules for the $G_i$'s (the experimentally measured
cross sections).
Discussing these subtleties will be the main
purpose of the rest of this paper. Here I only mention
that the OPE result is derived using dispersion relations
for the lepton-nucleon forward scattering amplitude and
in general there is always the possibility that the dispersion
relation does not converge or that the dispersion relation
requires a finite subtraction.

Similar results apply for the chirally odd parton distributions
$h^q_i(x)$ defined as \cite{jaji}
\begin{eqnarray}
& &\int \frac{d \lambda}{2\pi} e^{i\lambda x}
\langle PS|\bar{q}(0)\sigma^{\mu \nu}i\gamma_5q (\lambda n)|PS\rangle
\label{eq:h123}
\\
&=&2 \left[h^q_1(x)\left(S_T^\mu p^\nu-S_T^\nu p^\mu\right)/M \right.
\nonumber\\
& &+ h^q_L(x)M\left(p^\mu n^\nu - p^\nu n^\mu \right) (S \cdot n)
\nonumber\\
& &+ \left.
h^q_3(x) M\left(S_T^\mu n^\nu-S_T^\nu n^\mu\right) \right],
\nonumber
\end{eqnarray}
where $h^q_L(x)=h^q_1(x)+h^q_2(x)/2$.
Since they are chirally odd, they cannot be measured in
inclusive electro-production. One promising way of measuring them
is the polarized Drell-Yan process \cite{rhic} (see also Ref.\cite{ja92}):
If both target and beam are longitudinally polarized, one finds
for the spin assymetry\cite{sivers}
\begin{equation}
A_{LL} = \frac{\sum_q e_q^2 g_1^q(x) g^q_1(-y)}
{\sum_q e_q^2 f_1^q(x) f_1^{q}(-y)}
\label{eq:all}
{}.
\end{equation}
If both target and beam are transversally polarized, \cite{sivers}
\begin{equation}
A_{TT} = \frac{\sin^2 \theta \cos 2\phi}{1+\cos^2\theta}
\frac{\sum_q e_q^2 h_1^q(x) h^q_1(-y)}{\sum_q e_q^2 f_1^q(x) f_1^{q}(-y)}
\label{eq:att}
\end{equation}
and if one is longitudinal and the other transverse,
\cite{jaji,ralston}
\begin{eqnarray}
A_{LT} &=& \frac{2\sin 2\theta \cos \phi}{1+\cos^2\theta} \frac{M}{\sqrt{Q^2}}
\label{eq:alt}
\\
& &\frac{\sum_q e_q^2 \left( g_1^q(x) yg^q_T(-y) -xh_L^q(x) h^q_1(-y)\right)}
{\sum_q e_q^2 f_1^q(x) f_1^{q}(-y)}
\nonumber\\
{}.
\end{eqnarray}
An alternative way of measuring $h_1$ is through semi-inclusive
polarized electro-production \cite{jaji,hermes}.

Similarly to the sum rules for the $g^q_i$ one can derive
sum rules for the $h^q_i$ by integrating Eq.(\ref{eq:h123}) and one
finds \cite{nag92,mulders}, using again rotational invariance,
\begin{eqnarray}
\int_{-1}^1dx h^q_1(x)&=&\int_{-1}^1dx h^q_L(x)
\label{eq:hlsr}
\nonumber\\
\int_{-1}^1dx h^q_1(x)&=&2\int_{-1}^1dx h^q_3(x).
\label{eq:h3sr}
\end{eqnarray}
As is the case for the $g_3$-sum rule, the $h_3$-sum rule is most
certainly useless, but since measurements for $h_1$ and $h_L$ are
under way, it will soon be possible to test the $h_L$-sum rule (\ref{eq:hlsr})
experimentally.
In principle, the same comments that I made above for the
$g_T$ sum rule also apply for the $h_L$ sum rule.
However, as I will discuss in Section \ref{sec:pert}, it
seems to be violated already in perturbation theory.

The paper is organized as follows. In Section \ref{sec:ope},
I will explain
the connection between the light-cone correlation functions and
the experimentally measured quantities and
why the formal sum rules --- derived for the light-cone
correlation functions --- may be violated when applied to
the measured structure functions.
In Section \ref{sec:pert}, I will show that
the $h_2$ sum rule is violated in QCD.
In the following sections I will study some model
filed theories and simple (spin-independent)
higher-twist parton distributions and I will
attempt to shed some light on the physics of such
violations of sum rules.
In the Appendix, I will draw connections to the notorious
zero-modes plaguing the light-front community.

\section{Operator Product Expansion, Dispersion Relations and all that}
\label{sec:ope}
Deep inelastic scattering (DIS) experiments are performed in the
region $0<Q^2<2M\nu$ \cite{r1}. The operator product expansion is an
expansion around $Q^2=\infty$ and must diverge for  $Q^2<2M\nu$
because of the singularities of the hadronic tensor
(the physical discontinuity along the cut). The consequences of
these statements are easily demonstrated by considering
some hadronic tensor $T(x,Q^2)$, where $x = Q^2/2M\nu $.
To keep the discussion as general as possible,
we will consider here some generic forward scattering
amplitude, without specifying the current. This will allow us later to
draw very general conclusions. A very similar discussion,
for the specific example of $g_2(x)$, can be found in Ref.\cite{jacom}.

As an function of $x$, $T(x,Q^2)$ is analytic in the
complex plane, cut along $0<x<1$. The discontinuity
of the imaginary part along the cut is (up to trivial kinematic
factors) the experimentally
measured cross section, which we will denote by $G(x,Q^2)$.
In order to apply the OPE, one needs to relate $T(x,Q^2)$ for
$|x|>1$ to $T(x,Q^2)$ for $|x|<1$. From the theory of complex
functions it is well known that an analytic function is determined
by its singularities in the complex plane --- up to a
polynomial! This implies ($|x|>1$)
\begin{equation}
T(x,Q^2) = p\left(\frac{1}{x},Q^2\right) + \frac{1}{\pi}\int_0^1 dx^\prime
\frac{x}{x^2-{x^\prime}^2} G(x^\prime),
\label{eq:disp}
\end{equation}
where $p\left(\frac{1}{x},Q^2\right)$ is a polynomial in $\frac{1}{x}$ whose
coefficients do in general depend on $Q^2$.
In the language of Regge theory, such polynomial subtractions
are called ``fixed poles'' \cite{fixedpoles}.
Note that we specialized here on an odd function of $1/x$,
the generalization to the even case follows analogously.
Usually, only a few coefficients of the polynomial
are allowed to be nonzero, since otherwise unitarity bounds
are violated, but typically this places no restriction
on the lowest few terms.

For $x>1$ one can apply the OPE and one finds
\begin{equation}
T(x,Q^2) = \sum_{n=1,3,5,..}\frac{1}{x^n} a_n,
\end{equation}
where the coefficients are related to matrix elements of
local operators. If one allows for generalized functions, it is
always possible to find a function $g(x^\prime )$, such that
\begin{equation}
a_n=\int_{-1}^{1} dx^\prime {x^\prime}^{n-1} g(x^\prime)\, \, \quad \forall
n=1,3,...
\end{equation}
In the case of the moments appearing in the OPE for
DIS this function typically is a Fourier transform of a correlation
function along the light-cone, such as the $g_i$ and $h_i$
introduced in Eqs.(\ref{eq:g123},\ref{eq:h123}).

In order to demonstrate what happened if the subtraction polynomial
does not vanish, let us study an example, where
the polynomial is finite\footnote{Note that in general it may
happen that the integral in Eq.(\ref{eq:disp}) does not converge
and hence one needs an infinite subtraction. Such a case may
be considered as a limit of the situation assumed here.}
and first order\footnote{Typically, higher orders
are not allowed anyway because of unitarity constraints.}
$p \left( \frac{1}{x},Q^2 \right) = \frac{1}{x} p_1(Q^2)$.
Upon inserting $p$ in Eq.(\ref{eq:disp}) and introducing the
moments of the measured structure function
\begin{equation}
b_n \equiv \frac{1}{\pi} \int_0^1 dx^\prime {x^\prime}^{n-1}G(x^\prime)
\end{equation}
one thus finds
\begin{eqnarray}
a_n &=& b_n \quad \quad n=3,5,..\nonumber\\
a_1 &=& p_1+b_1 \, \, .
\label{eq:a1}
\end{eqnarray}
Such a result, if $p_1$ is nonzero, has a two major
consequences:
\begin{itemize}
\item Any sum rule, derived for $a_1$, fails to be satisfied
for $b_1$. Whether or not the sum rule derived using
the OPE applies also to the experimentally observed
structure functions depends on whether or not the
amplitudes satisfy un-subtracted dispersion relation.
\item Even when $G(x^\prime )$ is a smooth function, since
all moments except the lowest moment of $G(x)/\pi$ and $g(x)+g(-x)$
are the same, it must be that $g(x)$ contains a $\delta$-function
at the origin (which obviously afflicts only the lowest moment)
\begin{equation}
g(x)+g(-x) = p_1 \delta(x) + \frac{1}{\pi} G(x)
\label{eq:lowest}
\end{equation}
\end{itemize}
A few more comments are in order here. So far, I have only discussed
what happens if there is such a subtraction. In the rest of the
paper I will show (studying QCD perturbation theory as well
as some toy models) that this is actually a very common phenomenon.
In these examples I will furthermore demonstrate, that the
light-cone correlations indeed contain $\delta$-functions, even
when the "data" is very smooth, and that the coefficient
of the $\delta$-functions indeed coincides with subtraction
constants in dispersion relations.
Furthermore, I should emphasize that the $\delta$-functions
do {\bf not} arise from some obscure $Q^2 \rightarrow \infty$ limit
but from inverting a moment transformation at fixed and finite $Q^2$.

\section{Perturbative Analysis of \hbox{$h_L$} and
\hbox{$g_T$}}
\label{sec:pert}
So far, the discussion has been deliberately abstract.
Even though we have seen what happens {\it if} the
dispersion relations contain subtractions, we have not yet
shown {\it whether} this actually happens. As a first step in
this direction we will perform a perturbative analysis
of $h_L(x)$ and $g_T(x)$ in this section. Let us first look
at the ${\cal O}(g^2)$ corrections to $h_L(x)$
for a target that consists of a quark of momentum $P$ and
spin $S$, dressed by one gluon loop (Fig.\ref{fig:gluon}).
\begin{figure}
\unitlength1.cm
\begin{picture}(14,6)(2,-9.)
\includegraphics{gluon.ps}
\end{picture}
\caption{
One loop correction to
$g_T(x)$
and
$h_L(x)$ in QCD. The cross
represents the insertion of the light-cone correlator.}
\label{fig:gluon}
\end{figure}

For simplicity, we will assume $P_\perp=0$, but the generalization
to nonzero transverse momenta is straightforward.
{}From the definition (\ref{eq:h123}) one finds
\newpage
\noindent
\parbox[t]{18.cm}{
\begin{eqnarray}
h_L(x)\frac{S^-}{M} &=&-g^2\bar{u}(P,S) \int \frac{d^4k}{(2\pi)^4}
\delta(k^+-xp^+) \gamma^\mu
\frac{i}{\not \! \! \: k-m+i\varepsilon}
\sigma^{+-}i\gamma_5
\frac{i}{\not \! \! \: k-m+i\varepsilon}\gamma^\mu u(P,S)D_{\mu \nu}(P-k),
\end{eqnarray}
where
\begin{equation}
D_{\mu \nu}(q) = \frac{-i}{q^2+i\varepsilon}
\left( g^{\mu \nu} - \frac{q^\mu n^\nu+q^\nu n^\mu }{nq} \right)
\end{equation}

is the gluon propagator in the light-front gauge
\footnote{Since we are mostly concerned about the region
$x \approx 0$,
i.e. $k^+ \approx 0$, we don't have to worry about the
light-front singularities of the gluon propagator $D_{\mu \nu}(q)$,
which occur at $q^+=0$, i.e. $k^+ =P^+$.}
and $m$ is the current quark mass.
The term $\delta(k^+-xp^+)$ arises from the $\lambda$-integration
in Eq.(\ref{eq:h123}) which projects out quarks of
momentum fraction x.\\
Straightforward application of the $\gamma$-matrix algebra
yields (using $\not \! \! P u(P,S) = m\,u(P,S)$ and
dropping terms odd under
${\vec k}_\perp \rightarrow -{\vec k}_\perp$)}
\begin{eqnarray}
h_L(x)\frac{S^-}{M} &=&g^2\bar{u}(P,S) \int
\frac{d^4k}{(2\pi)^4}
\frac{\delta(k^+-xp^+)}{(p-k)^2+i\varepsilon}
\left\{
\frac{-4m\left(k^+\gamma^--k^-\gamma^+\right)i\gamma_5}
{(k^2-m^2+i\varepsilon )^2}
+\frac{1}{P^+-k^+}\frac{\left[ 2m\gamma^+i\gamma_5
+2k^+\sigma^{+-}i\gamma_5
\right]}{k^2-m^2+i\varepsilon } \right\}u(P,S).
\nonumber\\
& &\label{eq:numalg}
\end{eqnarray}
Most of the terms in Eq.(\ref{eq:numalg}) are harmless
as $k^+\rightarrow 0$. The only troublesome piece
arises from the term proportional to $k^-$.
In order to study its contribution further, one can make
use of the algebraic identity
\begin{equation}
k^- =P^- -\frac{({\vec P_\perp}-{\vec k_\perp})^2}{2(P^+-k^+)}
- \frac{(P-k)^2}{2(P^+-k^+)}.
\label{eq:alg}
\end{equation}
One can easily verify that the first two terms on r.h.s. of
Eq.(\ref{eq:alg}) are well behaved as $k^+\rightarrow 0$,
when inserted in Eq.(\ref{eq:numalg}).
The only troublemaker is the term proportional
to $(P-k)^2$ in the numerator. Using
\begin{equation}
\int \frac{dk^-}{2\pi} \frac{1}{(k^2-m^2+i\varepsilon)^2}
=\frac{i}{2} \frac{\delta(k^+)}{ {\vec k}_\perp^2 + m^2},
\end{equation}
one finds for the singular piece
\begin{eqnarray}
h_L^{sing}(x)MS^+
&=&  g^2 mS^+P^+
\int \frac{dk^+d^2k_\perp}{(2\pi)^3}
\frac{\delta(k^+-xP^+) \delta(k^+)}
{ {\vec k}_\perp^2 + m^2}
\nonumber\\
&=&g^2 \frac{mS^+}{\pi} \delta(x) \int \frac{d^2k_\perp}{(2\pi)^2}
\frac{1}{ {\vec k}_\perp^2 + m^2}
\nonumber\\
&=&g^2 \frac{mS^+}{4\pi^2} \delta(x) \log
\frac{\Lambda^2_\perp}{m^2}
\end{eqnarray}
where $\Lambda^2_\perp$ is some cutoff and where we used
$S \cdot P =0 $, i.e. $ S^-=-S^+M^2/2P^+$.
The factor $S^+$ on the r.h.s. arises from the matrix element
of $ i\gamma^+ \gamma_5$.
The precise numerical result is not so important here.
What is important is that straightforward evaluation of our perturbative model
yields a term proportional to a
$\delta$ function for $h_L(x)$.\\
\vspace{10.cm}\\
\begin{picture}(14,6)(0,-8.)
%\special{psfile=gn1.ps angle=-90 hscale=70 vscale=70}
\end{picture}
\vspace{7.cm}\\
\unitlength1.cm
\begin{picture}(5,0)(0,0)
\put(1.,0){\line(1,0){7.}}
\end{picture}\\
The singular term derived here is explicitly proportional
to the current quark mass, i.e. only for strange or heavier
quarks one expects a sizable contribution, and thus violation
of the $h_2$-sum rule. However, it is not known whether
this is only a lowest order artifact or whether this is a
more general result.
Certainly, using hard-soft factorization, this result implies
a $\delta (x)$ term with nonzero coefficient also in the
full (i.e. nonperturbative) result for $h_L(x)$ and
thus the $h_L$-sum rule should be violated for strange
and heavier quarks.

It should be emphasized that while we have
identified the quark mass as one source
of violation for the $h_L$-sum rule, it is
by no means clear whether or not there are
are other sources of further violation
which would also affect light $u$ and $d$ quarks.

Before we are going to investigate the physical origin
of the $\delta$-function, let us first take a look at
QCD perturbative corrections to $g_T(x)$ for a quark.
\newpage
\begin{eqnarray}
g_T(x)S^x_T &=&-g^2\bar{u}(P,S) \int \frac{d^4k}{(2\pi)^4}
\delta(k^+-xp^+) \gamma^\mu
\frac{i}{\not \! \! \: k-m+i\varepsilon}
\gamma^xi\gamma_5
\frac{i}{\not \! \! \: k-m+i\varepsilon}\gamma^\mu u(P,S)D_{\mu \nu}(P-k)
{}.
\end{eqnarray}
{}From the purely formal point-of-view the $\delta$-function
in $h_L(x)$ arises because of the term proportional to the
light-front energy $k^-\equiv (k^0-k^3)/\sqrt{2}$ in the
numerator, which cancels the energy dominator for the gluon
propagator. Armed with this insight, we can immediately
focus on the potentially dangerous terms in $g_T$
arising from the $k^-$ in the numerator of the fermion
propagator\footnote{One can easily verify that terms
with $k^-$ in the numerator from the gauge field propagator
are accompanied here by an explicit factor $\propto k^+$
in the fermion propagator and are thus harmless.}
\begin{eqnarray}
\gamma_\mu \left[ k^-\gamma^+\gamma^xi\gamma_5m
+ m \gamma^xi\gamma_5 k^-\gamma^+\right] \gamma^\mu
=0,
\end{eqnarray}
i.e. because of the Dirac algebra, the coefficient of
the potentially singular term in $g_T(x)$ vanishes.
In contrast to the result for $h_L(x)$,
at least in one loop order, radiative QCD corrections
do not give rise to $\delta$-functions in $g_T(x)$.

\section{Long-Range Correlations Along The Light-Cone in a Simple
Toy Model}
\label{sec:gn}
In the previous section, we have seen that sum rules
for higher-twist parton distributions can be easily
violated. However, so far, we lack an intuitive understanding
of the effect. In this (and the following)
section we will turn our attention to simple
toy models, where the same effect happens, but where it
is easier to analyse. For the same reason, we will
study the (spin independent) scalar higher-twist distribution
$e(x)$ in the rest of the paper. Both, working in 1+1 dimensions
and studying scalar distributions will simplify the algebra
involved considerably and thus allow us to shed some light
on the essential physics. The general conclusions should not
be affected by these simplifications.
One of the most simple examples for violations of sum rules for higher-twist
distributions is a $(1+1)$-dimensional model of the Gross--Neveu
type \cite{r10,r11}
\begin{equation}
{\cal L} = \bar{u}_i \left( i \not \!\partial - m_0\right) u_i + \bar{s}_i
\left( i \not \!\partial - m_0\right) s_i - {\lambda\over N_c} \bar{u}_i u_i
\bar{s}_j s_j\ \ ,
\end{equation}
where $i,j=1,...,N_C$ are ``color'' indices. Here we will restrict ourselves to
leading
order in $1/N_c$.  Suppose we probe a $u$-quark with an external vector
current which couples only to strange quarks.  From the hadronic
tensor
\begin{equation}
T^{\mu\nu}_{s/u} \left( q^2,p\cdot q\right) = i \int d^2x\, e^{iqx}
\langle
u,p\left| T \left( \bar{s} \gamma^\mu s(x) \bar{s}\gamma^\nu s(0)\right)
\right| u,p\rangle
\label{eq:2.1}
\end{equation}

\vspace*{.5cm}
\begin{picture}(5,0.25)(0,0)
\put(.5,0){\line(1,0){7.}}
\end{picture}\\[1.ex]

\noindent
we will consider only the symmetric part of the ``$+-$'' component.  A typical
Feynman diagram contributing to Eq.~(\ref{eq:2.1}) to
${\cal O}(N^0_c)$ (leading order)
is depicted in Fig.~\ref{fig:gn1}.
\begin{figure}
\unitlength1.cm
\begin{picture}(14,4)(0,-8.)
\includegraphics{gn1.ps}
\end{picture}
\caption{Leading order
$1/N_C$
contribution to
$T^{\mu\nu}_{s/u}$
in the
Gross--Neveu model.  The dotted lines indicate the flow of
momentum through
the external currents
$\bar{s}\gamma^\mu s$.
}
\label{fig:gn1}
\end{figure}
The OPE yields for $Q^2 = - q^2\to \infty$
\begin{equation}
T_{s/u} \equiv \nu\cdot {\left( T^{+-}_{s/u} +
T^{-+}_{s/u}\right) \over 2} = 2M^2 \int^1_{-1} dx' {x\over x^2 - {x'}^2}
e_{s/u} (x')
\label{eq:2.3}
\end{equation}
where
\begin{equation}
2M\,e_{s/u}(x) = \int^\infty_{-\infty} d\lambda\, e^{i\lambda x}
\langle
u,p\left| \bar{s}(0) s(\lambda n) \right| u,p\rangle
\ \ .\label{eq:2.4}
\end{equation}
Here $M$ is the constituent quark mass \cite{r10,r11}.  Clearly, $e_{s/u}(x)$
satisfies a ``sum rule'' (the {\it sigma-term sum rule}
\cite{r9})
\begin{equation}
\int^1_{-1} dx\, e_{s/u} (x) = {1\over 2M} \langle
u\left| \bar{s}s \right| u
\rangle\ \ ,
\label{eq:2.5}
\end{equation}
which is non-zero and ${\cal O}(N^0_c)$ \cite{r11}.  However, any attempt to
verify Eq.~(\ref{eq:2.5}) by studying physical
cross sections is doomed to fail, since
$\Im
T^{\mu\nu}_{s/u}=0$ to ${\cal O}(N^0_c)$.\footnote{This can be easily
verified by applying the cutting rules to Fig.~1.  Since $q^2<0$, at least one
of the cut lines will have a negative invariant mass.}

These seemingly contradictory results are consistent with dispersion theory.
The crucial point is that the OPE result (\ref{eq:2.3})
is derived in a kinematic
regime where $T^{\mu\nu}$ is purely real $|x|>1$ and which is not accessible
in DIS.  A relation between $T$ for $|x|>1$ (where $T$ is purely real) and
$\Im T$ for $|x|<1$ is established by means of a dispersion relation ($x>1$)
\begin{equation}
T_{s/u} (x) = p\left( {1\over x} \right) + {1\over \pi}\int^1_0 dx' {x\over
x^2 - {x'}^2} \Im T_{s/u} (x')\ \ ,\label{eq:2.6}
\end{equation}
where $p\left( 1/x\right)$ is an odd polynomial in $1/x$ which cannot be
determined from dispersion theory --- it corresponds to subtraction constants.
 If $p$ would vanish, then a comparison between Eqs.~(\ref{eq:2.6})
and (\ref{eq:2.3}) shows
\begin{equation}
{2\over \pi} \Im T_{s/u} (x') = 2M^2 \left[ e(x') + e(-x') \right]\ \
.\label{eq:2.7}\end{equation}
However, if for example, $p(1/x) = c/x$  then
\begin{equation}
{1\over \pi}\Im T_{s/u} (x') = c\cdot \delta(x') + 2M^2 \left[ e(x') +
e(-x') \right] \ \ .\label{eq:2.8}\end{equation}
Since $\Im T$, which is measured experimentally, does not contain
$\delta$-functions, we expect them to be contained in $e(x') + e(-x')$.  The
latter
will now be determined explicitly:  obviously
(Fig.~\ref{fig:gn2})
\begin{eqnarray}
& &\langle  u,p\left| \bar{s} (x) s(y) \right| u,p\rangle  =
\label{eq:2.9}\\
& &g^2 \int \frac{d^2k}{(2\pi)^2}
\bar{u}(p) u(p) tr \left( \frac{e^{ikx}}{\not \!k - m +i\epsilon} \
\frac{e^{-iky}}{ \not \!{k} - m+i\epsilon} \right)\ \
,\nonumber
\end{eqnarray}
{\it i.e.\/}
\begin{equation}
e_{s/u} (x') = g^2 \int {d\lambda \over 2\pi} e^{i\lambda x'} \int
{dk^+\,dk^-\over (2\pi)^2} \ {\left(k^2 + m^2\right) e^{-i{k^+\over
p^+}\cdot\lambda}\over
\left( 2k^+ k^- - m^2 +
i\epsilon\right)^2}   \ \ .
\label{eq:2.10}
\end{equation}
\begin{figure}
\unitlength1.cm
\begin{picture}(14,4.5)(1,-9.)
\includegraphics{gn2.ps}
\end{picture}
\caption{Leading order
$1/N_c$
Feynman diagram for
Eq.~(\ref{eq:2.9}).
}
\label{fig:gn2}
\end{figure}

Since the exponent in Eq.~(\ref{eq:2.10}) does not contain $k^-$,
one can always close
the contour in the complex $k^-$-plane such that no pole from the energy
denominators is enclosed.  For $k^+\not=0$ the surface term (from the
semi-circle in the complex $k^-$-plane) vanishes and one finds
\begin{equation}
\int dk^- {k^2 + m^2\over \left( 2k^+ k^- - m^2 + i\epsilon\right)^2}
\propto \delta (k^+)\ \
\label{eq:2.11}
\end{equation}
[The coefficient of proportionality is non-zero and calculable.  It turns out
to be such that $e_{s/u}(x)$ satisfies Eq.~(\ref{eq:2.5}), as one can verify by
comparison].  Therefore
\begin{equation}
e_{s/u}(x') \propto \delta(x')\ \ .\label{eq:2.12}
\end{equation}
Mathematically the $\delta$-function in Eq.(\ref{eq:2.12})
arises as a consequence of
long-range correlations along the light-cone.  In the Gross--Neveu model,
to leading order in $1/N_c$,
the light-cone correlator $\langle  u,p\left| \bar{s}(0) s(x^-) \right|
u,p\rangle $ is
independent of $x^-$ --- which is a consequence of Eq.~(2.11).  Physically,
this is related to the scalar coupling in the $t$-channel.  This remark
becomes more clear if we define the light-cone distributions by means of an
infinite momentum boost \cite{r12}.

Consider the equal time correlator ($p\equiv p^1$, $k\equiv k^1$)
\begin{equation}
\rho_p(k') = \int {dx_1\over 2\pi} e^{ikx_1} \langle  u,p\left| \bar{s}(0)
s(x_1) \right| u,p\rangle \ \ ,
\label{eq:2.13}
\end{equation}
which measures nothing but the conventional ({\it i.e.\/} not light-cone)
momentum distribution flowing through a strange current insertion (with scalar
quantum numbers).  It is related to the light-cone distribution through
\begin{equation}
e(x) = \lim\limits_{p\to\infty} p\cdot\rho_p (x\cdot p) \ \
.\label{eq:2.14}
\end{equation}
Now, as a consequence of the scalar $t$-channel coupling, $\rho_p(k')$ is
actually independent of $p$.  One finds
\begin{eqnarray}
\rho_p(k) &\propto& \int {dk^0\over 2\pi} tr\left( {\not \!{k} + m\over
k^2 - m^2 + i\epsilon}\right)^2 - {} ''m \to M'' \nonumber\\
&=& {{k}^2\over \left( {k}^2 +
m^2\right)^{3/2}} - {{k}^2\over \left( {k}^2 + M^2\right)^{3/2}}
\label{eq:2.15}
\end{eqnarray}
where we subtracted the contribution from a heavy regulator, in order to
render $\langle  u \left| \bar{s}s\right|u\rangle  = \int dk\,\rho_p(k)$
finite.
Clearly, $\rho_p(k)$ (Eq.~(\ref{eq:2.15})) is a localized function with finite
area.
Measuring all momenta in units of $p$ (which is after all what the area
preserving mapping in Eq.~(\ref{eq:2.14}) does) and performing the limit
$p\to\infty$
thus gives rise to a distribution of zero width but finite area --- a
representation for a $\delta$-function.\footnote{Using various test
functions can can verify that Eq.~(\ref{eq:2.14})
yields a ``pure'' $\delta$-function
and is not contaminated with higher multipoles, like $\delta''$.}

The fact that $\rho_p(k)$ is independent of $p$, to leading order in $1/N$ is
quite obvious, since the leading order in the $1/N$ approximation is similar
to a mean field approximation.  The expectation value of a scalar density,
taken in a plane wave state, is a constant --- independent of the momentum.
Therefore, the $u$-quark gives rise to a source for $s$-quarks [through the
point-like scalar
coupling in Eq.~(\ref{eq:2.1})] which also does not depend on its momentum
(\ref{eq:2.15}).

Thus one might suspect that the unique properties of the interaction term in
Eq,~(\ref{eq:2.1}) are the reason for our ``weird'' results:
\begin{itemize}
\item $\Im T$ smooth (actually identically zero in this example);
\item $\Im T$ violates the sum rule (the sum rule converges without
problems);
\item $e(x)$ contains a $\delta$-function; and
\item $e(x)$ satisfies the sum rule.
\end{itemize}
\noindent In particular, one might suspect that ``more realistic'' field
theories (like QCD$_{3+1}$) behave differently.  In the rest of this article
we will demonstrate that some of these properties are actually quite
generic for higher-twist distributions in many field theories (also
QCD$_{3+1}$).

A non-trivial and non-perturbative example has already been given in
Ref.~[13], where the twist-4 parton distribution
$f_4$ was analyzed in the context of QCD$_{1+1}$ ($N_c\to\infty$).
There, $\Im T^{--}$ (which scales against $f_4$ for non-zero $x$ in the
Bjorken limit) violates the sum rule which is derived for $f_4$.
Furthermore, $f_4$ contains a $\delta$-function, while $\lim\limits_{\rm Bj}\Im
T^{--}$ has only a mild (integrable) singularity for $x\to 0$.

In the above discussion, the presence of ultraviolet divergences in the
Gross--Neveu model forced us to introduce a regulator (cutoff).  In order to
keep the discussion simple the issue of renormalization (and elimination of
the cutoff dependence) has been avoided so far.  This will be done in the
following.  To leading order in $1/N_c$ one still finds $\Im T_{s/u} = 0$.
for the real part of the renormalized amplitude one finds
\cite{r10}, (Fig.\ref{fig:gn1})
\begin{eqnarray}
&&T^{\mu\nu}_{s/u} \left( q^2,\nu\right) = \label{eq:2.16}\\
&&g^2_{\sigma q\bar{q}} D_\sigma(0)
\int \frac{d^2k}{(2\pi)^2} tr \left[ \left(
\frac{1}{\not \!{k} -
M_F}\right)^2 \gamma^\mu \frac{1}{\not \!{k}+\not \!{q} - M_F} \gamma^\nu
\right]\nonumber\ \ ,
\end{eqnarray}
where $M_F$ is the physical fermion mass.  $g_{\sigma q\bar{q}}$ is the
physical quark-$\sigma$-meson coupling and $D_\sigma(0)$ is the
$\sigma$-meson propagator evaluated at $p^2=0$ (note that no momentum flows
through the $\sigma$-meson, since we are considering a forward amplitude).
Therefore,
\begin{eqnarray}
T_{s/u}\left( q^2,\nu\right) &=& \nu\left( T^{+-}_{s/u} + T^{-+}_{s/u} \right)
\nonumber\\
&=& c \frac{M^2_F\nu}{ Q^2} \int^1_0 dx
\frac{\left( Q^2\right)^2 x(1-x)}
{\left[Q^2x (1-x) + M^2_F\right]^2}
\nonumber\\
&=& c\cdot \frac{\nu\cdot M^2_F}{Q^2}\cdot
f\left( \frac{Q^2}{ M^2_F}\right)\ \
,
\label{eq:2.17}
\end{eqnarray}
where $c$ is some numerical constant.  If one now introduces a $Q^2$-dependent
parton distribution $e_{S/U}(x',Q^2)$ via the moment expansion
\begin{equation}
T_{s/u} (x,Q^2) = 2M^2_F \sum\limits_{\nu=0,2,4,\ldots} {1\over x^{\nu+1}}
\int^1_{-1} dx' {x'}^2 e_{s/u} (x',Q^2)
\label{eq:2.18}
\end{equation}
one finds the unique solution for the even part of $e_{s/u}(x',Q^2)$
\begin{eqnarray}
& &\frac{e_{s/u} \left( x', Q^2\right) + e_{s/u} \left( - x', Q^2\right)
}{2} =
\frac{c}{ 4M_F} \cdot \delta(x') \cdot f\left(
\frac{Q^2}{ M^2_F}\right)
\nonumber\\
& &\stackrel{Q^2\rightarrow\infty}{\longrightarrow}
\frac{c}{ 2M_F} \delta(x') \cdot \log \frac{Q^2}{M^2_F}
\ \ .
\label{eq:2.19}
\end{eqnarray}
Since $T^{\mu\nu}$ is independent of $\nu$, for all $Q^2$, one has
\hbox{$\nu\cdot
T\propto 1/x$}($x=Q^2/2M_F\nu$).  Therefore, the $\delta$-function is present
for all values of $Q^2$.  The coefficient multiplying $\delta(x')$ depends on
$\log Q^2/M^2_F$ which reflects the running of the coupling constant in the
Gross--Neveu model.

\section{Boosting slowly to infinite momentum}
\label{sec:inf}
In this Section we will continue our analysis of higher-twist distributions in
various field theories, although we will now use perturbation theory to the
lowest non-trivial order only.  Of
course, parton distributions in QCD cannot be calculated perturbatively, but
this is not what we attempt to do here.  The main motivation for this
perturbative study is to shed some light on possible mechanisms for the
violation of un-subtracted sum rules.  The use of perturbation theory will
allow us to explicitly compute various scattering amplitudes and correlations
analytically.  This is very important if one wants to understand the reason
for the failure of the sum rules mathematically as well as physically.

The example we will be studying at is again a $1+1$-dimensional model:
massive fermions coupled to massive bosons with a chirally invariant Yukawa
coupling (linear $\sigma$-model)
\begin{eqnarray}
{\cal L} = \bar{\psi}\left( i\not \! \partial - m\right) \psi &- \sigma \left(
\Box + m^2\right)\sigma - \pi \left(\Box + m^2\right) \pi\nonumber\\
&+\gamma\bar{\psi} \left( \sigma + i\gamma_5\pi\right) \psi\ \ ,
\label{eq:3.1}
\end{eqnarray}
where, for simplicity, all masses are taken to be equal.  The flavor structure
[which is suppressed in Eq.~(\ref{eq:3.1})] is assumed to be
such that only Fig.~\ref{fig:sigma}
contributes to order $\gamma^2$ to the hadronic tensor
\begin{equation}
T^{\mu\nu} \equiv
 i \int d^2x\, e^{iqx} \langle  u,p\left| T\left\{ \bar{d}\gamma^\mu
s(0) \bar{s} \gamma^\mu d(x)\right\} \right| u,p\rangle  \ \ .
\label{eq:3.2}
\end{equation}
This can, for example, be accomplished by having only the flavor combination
$\bar{u}d$ and $\bar{d}u$ couple to the mesons.  Straightforward application
of Feynman rules yields
\newpage
\noindent
\parbox[t]{18.cm}{
\begin{eqnarray}
\frac{1}{2} \left( T^{+-}+T^{-+}\right) &=& -4im^2 \gamma^2
\bar{u}(p) \int \frac{d^2k}{ (2\pi)^2}
\frac{\not \!{k}}{ \left( k^2 - m^2 +
i\epsilon\right)^2 \left( \left( k+q\right)^2 - m^2 + i\epsilon\right)
\left[ \left( p-k\right)^2 - m^2 + i\epsilon\right]} u(p) \nonumber\\
&=& \frac{16m^2\gamma^2}{ 4\pi} \int^1_0 dy \int^{1-y}_0 dz
\frac{\left( zm^2 -
y\,p\cdot q\right) \left( 1 - y - z\right) }{ \left[ y(1-y) q^2 + z(1-z)
p^2 +2yz\,p\cdot q - m^2 + i\epsilon\right]^3}\ \  .
\label{eq:3.3}
\end{eqnarray}
For the application of dispersion relations it is helpful to perform an
integration by parts in Eq.~(\ref{eq:3.3})
\begin{equation}
\frac{1}{2} \left( T^{+-}+T^{-+}\right) = \frac{m^2 \gamma^2}{ \pi}
\int^1_0 dy \frac{(1-y)}{\left[ y(1-y) q^2 - m^2 \right]^2}
+ \frac{m^2\gamma^2}{ \pi}\int^1_0 dy \int^{1-y}_0 dz \left\{ \frac{2m^2
(1-y-z)}{ D^3} + \frac{1}{ D^2}\right\}\ \ ,
\label{eq:3.4}
\end{equation}}
where
\begin{equation}
D = y(1-y) q^2 + z(1-z) p^2 + 2yz\, p\cdot q - m^2 + i\epsilon\ \ .
\label{eq:3.5}
\end{equation}
\begin{figure}
\unitlength1.cm
\begin{picture}(14,5)(1,-7.5)
\includegraphics{sigma.ps}
\end{picture}
\caption{Box diagram contribution to
$T^{\mu\nu}$
(\ref{eq:3.2})
in the linear
$\sigma$
-model
[Eq.~(\ref{eq:3.11})].
}
\label{fig:sigma}
\end{figure}
Most importantly, the first term in Eq.~(\ref{eq:3.4}),
which is purely real for
$q^2<0$, does not depend on $p\cdot q$.  The presence of this term thus ruins
the application of an un-subtracted dispersion relation to $\left(
T^{+-}+T^{-+}\right)/2$.  As explained in Section~\ref{sec:ope},
the presence of
subtraction constants in the dispersion relation is correlated with the
presence of a $\delta$-function in the light-cone momentum
distribution (\ref{eq:lowest}):

With
\begin{eqnarray}
&2\,e_{d/u}(x) = &
\label{eq:3.6}
\\ & &i\gamma^2  \int
\frac{d^2k}{ (2\pi)^2} \
\frac{\bar{u}(p)\not \!{k}\,\delta
\left(\frac{k^+}{p^+}-x\right)
 u(p)}{ \left( k^2 - m^2 +
i\epsilon\right)^2 \left( (p-k)^2 - m^2 + i\epsilon\right)}
\nonumber
\end{eqnarray}
and
\begin{eqnarray}
\not \!{k} &\equiv& k^+ \gamma^- + k^- \gamma^+ = k^+ \gamma^-
\label{eq:3.7}
\\
& &+ \left(
p^--{m^2\over 2\left( p^+-k^+\right)}\right)
 \gamma^+ - {(p-k)^2 - m^2 \over 2\left(
p^+-k^+\right)} \gamma^+
\nonumber
\end{eqnarray}
one finds
\vspace*{5cm}\\
\begin{picture}(5,0.25)(0,0)
\put(.5,0){\line(1,0){7.}}
\end{picture}\\[1.ex]
\begin{eqnarray}
2m\,e_{d/u} (x) &=& - 2\frac{m^2 \gamma^2}{ 4\pi}
\frac{\left( x + 1 -
{\frac{1}{1-x}} \right) (1-x) \theta(x) \theta(1-x) }{ \left[
m^2 x (1-x)-m^2\right]^2}
\nonumber\\
& & + \frac{2m\gamma^2}{ 4\pi} \
\frac{\delta(x) }{ m^2}
\nonumber\\
&=& \frac{\gamma^2}{ 2\pi m} \left\{ \frac{x^2 \theta(x) \theta(1-x)}{ \left[
x(1-x) - 1\right]^2]} - \delta(x) \right\}\ \ ,
\label{eq:3.8}
\end{eqnarray}
where the $\delta$-function came from the last term in
Eq.~(\ref{eq:3.5}).  It should
be emphasized here that the $\delta$-function in $e(x)$ is closely related to
the non-covariant counterterms in the Hamiltonian formulation of light-cone
field theories \cite{r14,mbrot,r16}. It should also be emphasized that the same
trick works in $3+1$-dimensions, {\it i.e.\/} there is nothing special about
two dimensions here.\\
\unitlength1.cm
Mathematically, the $\delta$-function appeared here because
\cite{r14,mbrot}
\begin{eqnarray}
\int \frac{dk^-}{2\pi} \ \frac{1}{ \left( k^2 - m^2+i\epsilon\right)^2} &=&
\delta
(k^+) \int \frac{d^2k}{2\pi} \ \frac{1}{ \left( k^2 - m^2 + i\epsilon\right)^2}
\nonumber\\ &=&
\frac{i}{2m^2} \delta(k^+)\ \ .
\label{eq:3.9}
\end{eqnarray}
{}From the physics point of view it is very instructive to consider a
careful boost to
infinite momentum.  In analogy to Eq.~(\ref{eq:2.13}) we consider the momentum
distributions of $d$-quarks, in the wave function of a dressed $u$-quark of
momentum $p$.  One finds for the interaction in Eq.~(\ref{eq:3.1})
\begin{eqnarray}
\rho_p(k)&=& \frac{m\gamma^2}{2\pi}\left\{ - \frac{1}{\left( k^2 +
m^2\right)^{3/2} }
\right.
\label{eq:3.10}
\\
& &\quad \quad + \left.\int^1_0 dy
\frac{(k-py)^2 + m^2 \left( y^2+\frac{1}{2}y-\frac{1}{2} \right) }
{\left[ (k-py)^2 - y(1-y)m^2 + m^2\right]^{5/2} }
\right\} \ .
\nonumber
\end{eqnarray}
Although there is no elementary scalar exchanged in the $t$-channel,
$\rho_p(k)$ contains a term that does not depend on $p$ at all.  Upon
substituting $k\to xp$ and performing the infinite momentum boost, this term
piles up to a $\delta$-function (again one verifies that no $\delta''$
components are present) as shown in Fig.~\ref{fig:rhosigma}.   Actually
$\lim\limits_{p\to\infty}
p\rho_p(xp)$ gives a result identical to the light-cone calculation
(\ref{eq:3.8}).
\begin{figure}
\unitlength1.cm
\begin{picture}(14,7.)(-13,2)
\includegraphics{rhosigma.ps}
\end{picture}
\caption{The momentum distribution of $d$-quarks in a moving $u$-quark
(\ref{eq:3.8})
for various values of the $u$-quark momentum.
$p\cdot \rho_p(x)$
is
plotted in units of
$\gamma^2/\pi m$.
}
\label{fig:rhosigma}
\end{figure}
Although the linear $\sigma$-model does not yield a point-like
scalar coupling in the t-channel, we observe a similar effect as in the
GN-model:
a part of the sea quark wave function does not follow the boosted
valence component. In coordinate space language this means that
only part of the sea quark distribution becomes Lorentz contracted
when the u-quark is boosted, while another component looks the
same in all frames! In 3+1 dimensional examples (similar to the one above)
this means that dressed particles do not necessarily become ``pancakes''
as $p\rightarrow \infty$ and some component of the wavefunction
retains a finite longitudinal extension. Whether that component
is ``visible'' in some experiment depends of course on the
quantum numbers of the probe and, in a manner which is not yet understood
completely, on the details of the underlying theory.

If one calculates only
\begin{equation}
\Im T \equiv {2pq\over \pi m} \Im {T^{+-}+T^{-+}\over 2}\ \ ,
\end{equation}
(Fig.~\ref{fig:imsigma}), one finds no indication whatsoever
about the presence of a $\delta$-function in $e(x)$.  For all values of $Q^2$,
$\Im T$ vanishes smoothly as $x_{\rm Bj} = Q^2/2p\cdot q\to 0$.\footnote{The
divergence for large $x$ in Fig.~\ref{fig:imsigma} is a purely
$1+1$-dimensional artifact and
arises from phase space factors.}  Also the $Q^2$ evolution does not rise to
any ``peak at small $x$.''
\begin{figure}
\unitlength1.cm
\begin{picture}(14,7)(.5,-8.7)
\includegraphics{imsigma.ps}
\end{picture}
\caption{
$p\cdot q \Im \left( T^{+-}+T^{-+}\right)\big/2$
(in units)
as a function of
$x=Q^2/2pq$
for various values of
$Q^2$.
Note the perfectly smooth behavior near
$x=0$
.}
\label{fig:imsigma}
\end{figure}
Furthermore, as $Q^2\to \infty$, $\Im T$
approaches $e(x)$ (Fig.~4) as long as $x\not=0$.  However, even though $\Im T$
is very smooth as $x\to 0$, the naive
version of
\begin{equation}
2m\int dx\, e_{d/u} (x) = \langle  u\left| \bar{d}d \right| u \rangle \ \
,\label{eq:3.11}
\end{equation}
where one replaces $2m\,e_{d/u} (x)$ by
fails.  This should be obvious since $\langle  u \left|\bar{d}d\right| u
\rangle $ is
negative to order $\gamma^2$ in the above model, while $\Im (T)$ is always
non-negative (from unitarity).

A different perspective is reached by computing the real part of $T$ as well.
 In the Bjorken limit one finds
\begin{eqnarray}
\Im T(x) &=& \frac{\gamma^2}{2} x^2 \frac{\theta(x)\theta(1-x)}{
\left[ x(1-x) - 1\right]^2} \nonumber\\
\Re T(x) &=& \frac{1}{\pi} \int^1_0 dx'\frac{\Im T(x')}{x-x'} -
\frac{\gamma^2}{2\pi x} \qquad (x>1)
\label{eq:3.13}
\end{eqnarray}
which is consistent with the result from the OPE.\footnote{Notice that,
since the currents in Eq.~(3.2) are not charge conjugation eigenstates, $e(x)$
and $T(x)$ do not have a definite transformation property under $x\to -x$
either.}
\begin{equation}
\Re T(x) = m\int^1_{-1} dx' {e_{d/u}(x') \over x-x'}\ \ .
\label{eq:3.14}
\end{equation}
However, the presence of the subtraction term in
(\ref{eq:3.13})
means that one can
identify $\Im T(x')$ with $e(x')$ only for $x'\not=0$.

It is instructive to re-analyze the situation from the point of view of the
infinite momentum frame.\footnote{Of course, by making $p$ in Eq.~(3.8)
large, one boosts the target and not the observer.  However, this is
equivalent.}  As $p\to\infty$ one almost recovers the naive picture, where all
virtual constituents carry a positive, finite fraction of the hadrons'
momentum: $p\rho(xp)$ drops sharply at $x=1$ for $p=10^3$ in Fig.~4 and
vanishes for $x<0$.  This is consistent with the absence of vacuum
fluctuations in the $\infty$-momentum frame because this would require at
least one particle carrying a negative momentum fraction \cite{r12}.  However,
some component of the wave function of the $u$-quark ``lacks behind'' in the
boost (\ref{eq:3.10}).  This can be summarized in the following physical
picture.  As
$p\to\infty$ most of the soft modes decouple from the system, which
is partly responsible for the simplified dynamics in the
{\it infinite momentum frame}. Of course these modes don't completely disappear
but get
concentrated in the region near $x=0$.  While leading twist distributions
couple only weakly to this
component of the wave function, there is a stronger coupling of the
higher-twist distributions.  That is why leading twist distributions do not
contain these $\delta$-functions.  A very similar and related effect was first
observed in the context of QED$_{3+1}$ \cite{r19}.  There it was shown that
certain
connected vacuum graphs do survive the $p\to\infty$ limit.  Since $e(x)$ is,
up to the momentum projection, proportional to a mass insertion, it should be
clear that the non-covariant piece in the self-energy of an electron in QED
--- which is all that survives from the $z$-graph as $p\to\infty$ --- is (up
to mass derivative) proportional to the $\delta$-function contribution one
finds for $e(x)$ in QED.

I should emphasize again that, if such a situation
applies to experimentally measured higher twist distributions, one would not be
able to ``see''
any hint about the singular behavior of the
light-cone distribution by just looking at the
structure function.

\section{Summary}
We have investigated higher-twist distributions in a variety of field theories.
 This included perturbative examples in $1+1$ as well and $3+1$
dimensions.\footnote{For a non-perturbative example in $1+1$ dimensions,
see Ref.~\cite{r13} for a discussion of $f_4$ in the context of QCD$_{1+1}$
($N_c\to\infty$). See also
Ref.~\cite{mbsg} for a discussion of the scalar
density in the sine-Gordon model.}  In most cases the naive sum rules for the
lowest moments
of higher-twist distributions are violated when summed over the
``experimentally determined''\footnote{Actually, the imaginary part of
some scattering amplitudes.  Of course, there can only be Gedanken experiments
in model field theories.} distributions.  For non-zero $x_{\rm Bj}$, these
structure functions scale towards the Fourier transformed quark-quark
correlation in the hadron along the light-cone.  The phenomenon which causes
the failure of the sum rules for higher-twist structure functions is a
$\delta$-function at the origin in the light-cone distribution.  Structure
function and light-cone distribution coincide for non-zero $x$, while the
$\delta$-function is absent in the structure function.  The sum-rules are
valid when one includes the $\delta$-function, {\it i.e.\/} when applied to
the light-cone distribution.  The structure functions then cease to satisfy the
sum rule because they do not contain the $\delta$-function.  In that sense the
$\delta$-function destroys the sum rule.  From a theorist's point of view, the
$\delta$-function restores it, {\it i.e.\/} when added to the structure
function the combined result should reproduce the sum rule.  The last point
may however be of purely academic interest, since one cannot measure the
light-cone correlation directly.

The relation between light-cone correlations and structure functions is usually
based on the operator product expansion combined with dispersion relations.
Unless one has a theoretical or experimental constraints on the high-energy
behavior of the real part of the scattering amplitude, subtraction constants
cannot be ruled out.  In fact, the coefficient of the $\delta$-function is
proportional to such a high-energy subtraction. This also shows that the
long-range correlation along the light-cone are consistent with the OPE.

In coordinate space language this means that only part of the sea quark
distribution gets Lorentz contracted when the $u$-quark is boosted
(giving rise to the finite-$x_{Bj}$ component), while
another component looks the same in all frames (contributing only at
$x_{Bj}=0$).  In $3+1$-dimensional examples
(similar to the one above) this means that the hadron does not become a
``pancake'' as $p\to\infty$.  Only part of the wave function is Lorentz
contracted while some rest retains its spherical shape.

There is nothing wrong or inconsistent about this result.  It is only another
example where naive intuition (in this case based on the Lorentz
transformation properties in free field theory) fails.  Interacting fields may
have very complicated transformation properties which, in the above example,
cause that part of the wavefunction does not Lorentz contract in the same way
as free fields do.  This also ``explains'' why higher-twist distributions are
more vulnerable to these effects, since, in the parton model, they contain
interaction terms explicitly.

{}From the point of view of null plane quantization, the $\delta$-functions are
a manifestation of zero-modes \cite{le:ap,kent,ho:vac,r21,r22,alex,bret} in the
hadronic wave function (see
Appendix~A).  Higher-twist parton distributions are defined through
correlations which involve bad currents ({\it i.e.\/} $\psi^{(-)}$ component
of the spinors).  When solving the constraint equation for $\psi^{(-)}$ one
has to specify boundary conditions at $x^-=\pm\infty$.  Non-zero values at the
boundary are of course related to $\delta$-functions in the Fourier transform.
 Another (not necessarily independent) connection to the failure of canonical
light-cone quantization (without zero modes) is observed by considering the
singular part of $e(x)$.  In perturbation theory one finds that the derivative
of the non-covariant $\gamma^+/p^+$ term in the fermion self-energy with
respect to the quark mass is proportional to the $\delta$-function coefficient
in $e(x)$.  Again this should not be a surprise, since those terms are caused
by an improper treatment of the zero modes \cite{mbrot} in canonical light-cone
perturbation theory \cite{r23} --- unless one uses several Pauli--Villars
regulators (more than in a manifestly covariant approach)\cite{mbrot,r24}.

In practical applications, most sum rules for higher-twist distributions are
probably useless since one expects them to diverge
\cite{r25,r26}.  The discussion of
$e(x)$ in this work should therefore be considered only a pedagogical example.
 One exception seems to be $g_2$.  From a Regge pole analysis one expects the
Burkhardt--Cottingham sum rule to converge \cite{bc}.  Also, as
shown in Section~\ref{sec:pert},
perturbation theory does not indicate any divergences in QCD.  Furthermore,
Regge pole analysis \cite{bc} as well as perturbative considerations
(Section~IV)
suggest validity of the sum rule. However, no strict predictions for $g_2$
have been made on a non-perturbative basis so far and
at this point one has to wait for the experimental result
to see whether the $g_2$-sum rule is valid.

For strange or heavier quarks, the $h_L$ sum rule seems to be most
certainly violated. The perturbative analysis in this work
only yielded a violation proportional to the quark mass and thus
one might be tempted to assume validity of the $h_L$ sum rule
for $u$ and $d$ quarks. However, it is not clear whether other
effects, which have not been considered in this work, lead to a violation
for light quarks as well.

Even if these sum-rules are violated, the OPE remains valid.
What needs to be added is a subtraction in the dispersion relation.
Note that such a mechanism cannot be used to ``explain'' the
spin crisis, since $g_1$ is leading twist and thus a
subtraction in the dispersion relation would be in conflict with
unitarity constraints.
\appendix
\section{\protect{$e(x)$} ON A FINITE INTERVAL}
The $\delta$-function at the origin in $e(x)$ strongly suggests a relation to
zero modes in light-cone quantization \cite{le:ap,r21,r22,alex,bret}.
Due to the severe infrared
singularities in light-cone quantization (the kinetic energy diverges as
$k^+\to 0$) it turns out to be necessary to go on a finite interval.
Furthermore, following Lenz {\it et al.\/}\cite{le:ap}, we will use a tilted
coordinate system
\begin{equation}
\hat x^- = x^-\ \ ,\qquad \hat x^+ = x^++\epsilon{x^-\over L}\ \
,\label{eq:A.1}
\end{equation}
where the $\hat{\phantom x}$ refers to the ``usual'' light-cone coordinates
\begin{equation}
\hat x^+ = \left( x^0\pm x^1\right)\big/\sqrt{2}\ \ .
\label{eq:A.2}
\end{equation}
The notation in this Appendix will thus be different from the rest of the
paper in order to allow for a direct comparison with Ref.~\cite{le:ap}, which
is
strongly recommended to supplement this Appendix.

This implies for example for scalar products
\begin{equation}
a\cdot b = \hat a_+ \hat b_- + \hat a_- \hat b_+ = a_+ \left( b_- +
{\epsilon\over L} b_+\right) + b_+ \left( a_- + {\epsilon\over L}a_+\right)\ \
.\label{eq:A.3}
\end{equation}
$L$ refers to the length of the interval and we will impose periodic boundary
conditions in $x^-$.  The canonically conjugate momentum $p_-$ thus becomes
quantized
\begin{equation}
p_{-n} = {2n\,\pi\over L}\ \ .
\label{eq:A.4}
\end{equation}
For the linear $\sigma$-model (\ref{eq:3.1}) one thus finds to order $g^2$
\begin{eqnarray}
e(k_-) &\propto& \int dk_+ \frac{\not \!{k} }{\left( k^2 - m^2 +
i\epsilon\right)^2} \  \frac{1}{(p-k)^2 - m^2 + i\epsilon}
\nonumber\\
&=& \frac{c}{ \sqrt{\epsilon\cdot L}} \left\{ \int^1_0 dx\left[ \frac{1}{
2D^{3/2}} - \frac{3}{ 4}\  \frac{m^2(1-x)}{ D^{5/2}} \right]
\right.\nonumber\\
& &\quad \quad \quad \quad \quad \quad  - \left.
\frac{1}{ 2}\
\left[ {\frac{L}{2\epsilon}} k^2_-+ m^2\right]^{-3/2} \right\}\
,
\label{eq:A.6}
\end{eqnarray}
where
\begin{equation}
D = {L\over 2\epsilon} \left( k_- - xp_-\right)^2 - x(1-x) m^2 + m^2\ \
.\label{eq:A.7}
\end{equation}
Of course $k_-$ as well as $p_-$ take on only discrete values (A.4).  Here $c$
is some numerical constant proportional to $g^2$.  One immediately recognizes
the similarity to Eq.~(3.8), which is not completely accidental, since the
infinite momentum boost also gives some kind of regularized representation for
$e(x)$.  The canonical light-cone limit for $e(k_-)$ is obtained by taking
$\epsilon L m^2\to 0$ while keeping $n_k$ and $n_p$ fixed ($k_- = 2\pi n_k/L$,
$p_- = 2\pi n_p/L$).  Using
\begin{equation}
\int^1_0 {dx\over \sqrt{\epsilon L}} f\left( {x-y\over \sqrt{\epsilon
L}}\right) \stackrel{\sqrt{\varepsilon L}\rightarrow 0}{\longrightarrow}
\int^\infty_{-\infty} dx\, f(x)\ \ ,
\end{equation}
this yields ($y = {k_-\over p_-}$),
\begin{equation}
e(k_-) \to c\cdot \sqrt{2} {y^2 m^2\over \left[ y(1-y)m^2 - m^2\right]^2}
\theta(y) \theta(1-y)\ \ ,
 \label{eq:A.8}
\end{equation}
for $n_k\not=0$.  However, for $n_k=0$,
\begin{equation}
e(0) \to - {c\over \sqrt{\epsilon L}} \cdot {1\over m^3}\ \
,\label{eq:A.9}
\end{equation}
{\it i.e.\/} the zero mode contribution diverges.

The mere fact that $e(0)$ diverges reflects the existence of the
$\delta$-function in the continuum calculation.  However, if one actually
wants to make a quantitative comparison with Section~III, one has to perform
a more sophisticated continuum limit where also the hadron momentum $n_k$ (in
discrete units) approaches infinity.

\end{document}